\documentclass[12pt,preprint]{elsarticle}
\usepackage[latin9]{inputenc}
\usepackage{units}
\usepackage{textcomp}
\usepackage{amsmath}
\usepackage{amssymb}
\usepackage{graphicx}
\PassOptionsToPackage{normalem}{ulem}
\usepackage{ulem}

\makeatletter

\newcommand{\lyxmathsym}[1]{\ifmmode\begingroup\def\b@ld{bold}
  \text{\ifx\math@version\b@ld\bfseries\fi#1}\endgroup\else#1\fi}

\newcommand{\lyxdot}{.}

\usepackage{a4wide}

\makeatother

\begin{document}

\title{PHAST: Protein-like Heteropolymer Analysis by Statistical Thermodynamics}

\author{Rafael B. Frigori}

\ead{frigori@utfpr.edu.br}

\address{Universidade Tecnológica Federal do Paraná, Rua Cristo Rei 19, CEP
85902-490, Toledo (PR), Brazil }
\begin{abstract}
PHAST is a software package written in standard Fortran, with MPI
and CUDA extensions, able to efficiently perform parallel multicanonical
Monte Carlo simulations of single or multiple heteropolymeric chains,
as coarse-grained models for proteins. The outcome data can be straightforwardly
analyzed within its microcanonical Statistical Thermodynamics module,
which allows for computing the entropy, caloric curve, specific heat
and free energies. As a case study, we investigate the aggregation
of heteropolymers bioinspired on $A\beta_{25-33}$ fragments and their
cross-seeding with $IAPP_{20-29}$ isoforms. Excellent parallel scaling
is observed, even under numerically difficult first-order like phase
transitions, which are properly described by the built-in fully reconfigurable
force fields. Still, the package is free and open source, this shall
motivate users to readily adapt it to specific purposes.
\end{abstract}
\maketitle

\section*{PROGRAM SUMMARY}

\begin{footnotesize} 
\begin{description}
\item [{\em Authors:}] Rafael Bertolini Frigori
\item [{\em Program Title:}] PHAST
\item [{\em Journal Reference:}] 

\item [{\em Catalogue identifier:}] 

\item [{\em Licensing provisions:}] GNU/GPL version 3
\item [{\em Programming language:}] FORTRAN 90, MPICH 3.0.4, CUDA 8.0
\item [{\em Computer:}] PC 
\item [{\em Operating system:}] GNU/Linux (3.13.0-46), should also work
on any Unix-based operational system. 
\item [{\em Supplementary material:}]~
\item [{\em Keywords:}] Folding: thermodynamics, statistical mechanics,
models, and pathways; Biopolymers, biopolymerization; Monte Carlo
methods; Thermodynamics and statistical mechanics


\item [{\em PACS:}] 87.15.Cc; 82.35.Pq; 05.10.Ln; 64.70.qd
\item [{\em Classification:}]~
\item [{\em External routines/libraries:}] cuRAND (CUDA), Grace-5.1.22
or higher
\item [{\em Subprograms used:}]~
\item [{\em Nature of the problem:}] Nowadays powerful multicore processors
(CPUs) and Graphical Processing Units (GPUs) became much popular and
cost-effective, so enabling thermostatistical studies of complex molecular
systems to be performed even in personal computers. PHAST provides
not only an easily-reconfigurable parallel program for Monte Carlo
simulations of linear heteropolymers, as coarse-grained models of
proteins, but also permits the automatized microcanonical thermodynamic
analysis of those systems.
\item [{\em Solution method:}] PHAST has three main modules for complementary
tasks as: to map any .pdb file-sequence to its inner lexicon by using
a configurable hydrophobic scale, while the main simulational module
performs parallel Monte Carlo simulations in the multicanonical (MUCA)
ensemble and, the analysis module which extracts microcanonical observables
from MUCA weights.
\end{description}
\end{footnotesize}

\section{Introduction}

Linear heteropolymeric chains of amino acids are known as polypeptides.
Proteins, on the other hand, are a large class of biological polymers
containing at least one long polypeptide. Together, they constitute
a set of macromolecules performing a vast array of functions within
organisms, whose specificities strongly depend on their detailed intramolecular
interactions that leads to three-dimensional (native) structures by
a folding mechanism. The thermodynamic hypothesis \cite{Anfisen}
states that the native structure of a protein is a unique, stable
and kinetically accessible minimum of the free energy solely determined
by its (primary) sequence of amino acids. However, eventual misfolding
may produce dysfunctional proteins, so inducing the formation of cytotoxic
amyloid aggregates, which can culminate on degenerative diseases as
type 2 Diabetes Mellitus (DM2) \cite{Diabetes_2} or Alzheimer (AD)
\cite{Alzheimer}.

Once those biopolymers are typical representatives of the so-called
small-systems, where in general the equivalence of statistical ensembles
does not hold (see \cite{Ensemble_inequivalence} and references
therein), the direct computation of their density of states is a well
suited investigation method to result on the system microcanonical
thermodynamics \cite{Gross_microcanonical_thermostatistics}. Nevertheless,
performing such calculations at a high enough accuracy implies on
accumulating large amounts of statistical data through Monte Carlo
methods, as Wang-Landau \cite{Wang_Landau} or the multicanonical
(MUCA) ensemble \cite{BERG_MUCA}. Thus, it consists of a huge computational
challenge, which can be reasonably alleviated just by conjugating
efficient parallel algorithms and coarse-grained molecular force fields.

To accomplish these demands we designed PHAST, a simple and easy-to-use
open source package that enables users to simulate and analyze the
microcanonical thermostatistics of general models for semi-flexible
linear polymers \cite{FENE_SINGLE-chain_phase_diagram,Long_review_Janke},
as coarse-grained proteins \cite{AB_model,Microcanonical_analysis_Peptide_Aggregation_Janke-PRL,Microcan_thermostatistics_coarse-grained_proteins,h-c-r-IAPP_agg_PRE_RBFrigori}.
The program brings fully reconfigurable built-in force fields that
can be readily modified and extended, this makes it specially apt
for prototyping of many-body interacting systems, as polymers embedded
in crowded solutions \cite{Macromolecular_crowding}. It is written
in standard Fortran and possess MPI and CUDA extensions \cite{MPI,CUDA},
so being able to efficiently run on most modern parallel computer
architectures. The distribution is forbidden for commercial or military
purposes, but we hope that PHAST is useful and could be eventually
incorporated in other programs, while we kindly request to be informed
accordingly. Still, it is naturally supposed to be acknowledged in
all resulting publications.

The article is organized as follows, in the Section 2 we review some
simulation background, this includes coarse-grained modelling and
the derivation of microcanonical thermostatistics from parallel multicanonical
simulations. The Section 3 encompasses an in-dept overview of application
modules. In the Section 4 two concrete case studies involving heteropolymers
bioinspired on amyloidogenic proteins are presented, so the aggregation
of $A\beta_{25-33}$ fragments and their further cross-seeding with
$IAPP_{20-29}$ isoforms are investigated in the context of a minimal
coarse-grained model. The parallel scaling of such simulations is
rigorously addressed, while their physical significance is outlined,
when possible, by comparisons to other results from the literature.
The Section 5 brings our conclusions and perspectives for further
developments.

\section{Simulation background}

\subsection{Modelling proteins and polymers}

Coarse-graining is a widely employed modelling strategy to reduce
the degrees of freedom of many-body macromolecular systems as heteropolymers
\cite{Long_review_Janke,Coarse_grained_models_proteins}. Moreover,
the study of protein folding and aggregation, specially in crowded
media \cite{Protein_crowding_effects}, has also largely benefited
from such approach \cite{Coarse_grained_models_proteins}. In this
vein, PHAST incorporates a configurable force field able to describe
semi-flexible linear heteropolymers%
\footnote{While it is a considerable limitation, once in chemically relevant
situations side chains may become crucial, this implementation is
computationally quite effective.%
} \cite{FENE_SINGLE-chain_phase_diagram,Agg_theta-polymers,Micro_Can_Analysis_Homopolymer,Multicanonical_study_models_folding_heteropolymers}
and general properties of protein-like structural phase transitions
\cite{AB_model,Microcanonical_analysis_Peptide_Aggregation_Janke-PRL,Microcan_thermostatistics_coarse-grained_proteins,h-c-r-IAPP_agg_PRE_RBFrigori}.
In the present model, a primary sequence is translated to a simpler
one according to a predefined rule (i.e. sorting each residue by its
hydrophobic character), so the amino acids on a proteic backbone are
mapped to a coarser chain of pseudo-atoms (beads).

Once original molecular interactions are mapped into simpler ones,
non-bonded interactions among beads are represented through effective
hydrophobic forces \cite{Hydrophobic_force}, as given by a Lennard-Jones
potential $\left(LJ\right)$. Thus, it is natural to adopt a specific
hydrophobic scale \cite{Roseman_scale} to perform the aforesaid
sequence translation, see Figure 1. By considering that hydrophobic
residues $\left(A\right)$ form more strongly-interactive cores than
hydrophilic ones $\left(B\right),$ the coupling constant for pairs
of nonadjacent beads $\left(\sigma_{i},\sigma_{j}\right)$ may be
defined by 
\begin{equation}
C_{LJ}\left(\sigma_{i},\sigma_{j}\right)=\left\{ \begin{array}{cc}
+1 & \sigma_{i}=\sigma_{j}=A\\
+\nicefrac{1}{2} & \sigma_{i}=\sigma_{j}=B\\
-\nicefrac{1}{2} & \sigma_{i}\neq\sigma_{j}
\end{array}\right.\label{Hydrophobicity_constant}
\end{equation}

Notwithstanding the fact that the protein backbones are rather stiff,
they may be well modelled as bead-spring heteropolymers. Therefore,
the bounds between successive beads located at a distance $r_{l,l+1}$
are described by the usual finitely extensible nonlinear elastic potential
(FENE) \cite{FENE_potential}, while a bending potential term between
three successive beads is considered as proportional to $\left(1-\cos\theta_{k}\right).$
The resulting energy for a single chain composed by $N$ beads may
be written as

\begin{figure}
\begin{centering}
\includegraphics[clip,height=7cm]{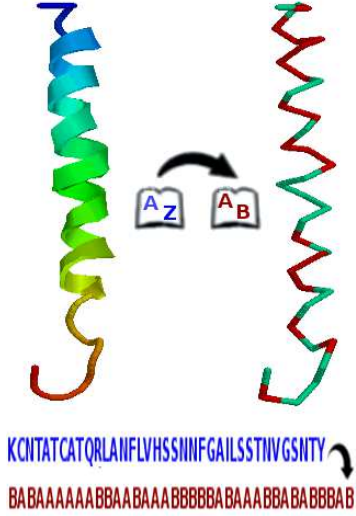}
\par\end{centering}

\caption{A chosen hydrophobicity scale is employed to map the FASTA sequence
(A--Z) of a protein (on the left) extracted from the Protein Data
Bank to another sequence, expressed by a two-letter code (A--B), which
is ultimately used to build the coarse-grained one-bead protein model
(on the right).}
\end{figure}

\begin{equation}
\begin{array}{c}
H_{single}=\kappa_{F}R^{2}\sum_{l=1}^{N-1}C_{F}\left(\sigma_{l},\sigma_{l+1}\right)\ln\left(1-\left[\left(r_{l,l+1}-r_{0}\right)/R\right]^{2}\right)+\\
\\
\kappa_{C}\sum_{k=1}^{N-2}\left(1-\cos\theta_{k}\right)+\kappa_{LJ_{INTRA}}\sum_{i=1}^{N-2}\sum_{j=1+1}^{N}\left[r_{ij}^{-12}-C_{LJ}\left(\sigma_{i},\sigma_{j}\right)r_{ij}^{-6}\right],
\end{array}\label{H_one_protein}
\end{equation}
where the first term (FENE potential) has $\kappa_{F},$ $r_{0}$
and $R$ as general coupling constants for homopolymers, while $C_{F}\left(\sigma_{l},\sigma_{l+1}\right)$
allows for introducing different spring stiffness for heteropolymers.
Moreover, $\kappa_{C}$ sets the chain rigidity against bends by an
angle $\theta_{k},$ while $\kappa_{LJ_{INTRA}}$ defines the coupling
constant of the non-bonded  intramolecular $LJ$ potential between
pairs of beads $\left(i,j\right)$ at a distance $r_{ij}$.

Finally, $M$ different chains are made to constitute a many-body
interacting system by the following multi-chain potential 

\begin{equation}
H_{multi}=\sum_{k=1}^{M}\left(H_{single,k}+\sum_{l>k}\sum_{i,j=1}^{N}\kappa_{LJ_{INTER}}\left[r_{l_{i}k_{j}}^{-12}-C_{LJ}\left(\sigma_{l_{i}},\sigma_{k_{j}}\right)r_{l_{i}k_{j}}^{-6}\right]\right),\label{H_multi_protein}
\end{equation}
where $H_{single,k}$ is given by Eq.(\ref{H_one_protein}), and the
second term describes the non-bonded intermolecular $LJ$ potential
between pairs of beads $\left(l_{i},k_{j}\right)$ at a distance $r_{l_{i}k_{j}},$
whose coupling constant is $\kappa_{LJ_{INTER}}.$

\subsection{Parallel multicanonical simulations}

It has been shown that microcanonical analysis of Monte Carlo simulations
provide a powerful tool not only to precisely characterize structural
phase transitions on small-systems \cite{Gross_microcanonical_thermostatistics,FENE_SINGLE-chain_phase_diagram},
where ensemble equivalence does not hold (see \cite{Ensemble_inequivalence},
and references therein), but also allows for neatly distinguish between
good and bad folders among protein-like heteropolymers \cite{Microcan_vs_Canonical_analysis}.
Most of straightforward implementations of such methodology rely on
the accurate determination of systems density of states $\Omega\left(E\right),$
which may be achieved by the Wang-Landau method \cite{Wang_Landau},
or with generalized ensembles as the multicanonical \cite{BERG_MUCA}.

Once knowing $\Omega\left(E\right),$ the microcanonical thermostatistics
can be established by the usual Boltzmann formula for the entropy
$S\left(E\right),$ as follows 
\begin{equation}
S\left(E\right)=k_{B}\ln\Omega\left(E\right).\label{Boltzmann_formula}
\end{equation}
This leads to the microcanonical connection to thermodynamical observables
\cite{Gross_microcanonical_thermostatistics}. For instance, numerical
derivatives of the entropy $S\left(E\right)$ can be taken to compute
quantities of interest, as the microcanonical caloric curve, relating
the temperature $T$ to the internal energy $E,$ given by
\begin{equation}
\, k_{B}\beta\left(E\right)\equiv T\left(E\right)^{-1}=\frac{\partial S}{\partial E}.\label{Caloric_curve}
\end{equation}
Also, the microcanonical specific heat $C_{V}\left(E\right)$ is expressed
as
\begin{equation}
C_{V}\left(E\right)=\frac{dE}{dT}=-\left(\frac{\partial S}{\partial E}\right)^{2}\left(\frac{\partial^{2}S}{\partial E^{2}}\right)^{-1},\label{Specific_heat}
\end{equation}
while the free energy $F\left(E\right),$ at a fixed temperature $T_{c},$
is computed by 
\begin{equation}
F\left(E\right)=E-\left.\left(\frac{\partial S}{\partial E}\right)^{-1}\right|_{E=E\left(T_{c}\right)}S\left(E\right).\label{Free_energy}
\end{equation}

As previously stated, the multicanonical ensemble offers a quite practical
method to estimate $\Omega\left(E\right)$ through Monte Carlo simulations,
and so, to obtain the microcanonical thermostatistics of any physical
system. Basically, on this approach the entropy is written as a piecewise
function of the discretized energies $E_{k}$ and a set of MUCA coefficients
$\left\{ \beta_{k},\alpha_{k}\right\} ,$ as follows
\begin{equation}
S_{muca}(E_{k})=\beta_{k}E_{k}\lyxmathsym{\textminus}\alpha_{k}.\label{MUCA_entropy}
\end{equation}
The Monte Carlo simulations in the multicanonical ensemble naturally
employ the so-called generalized MUCA weights $\,\omega_{muca}\left(E_{k}\right)\propto e^{-\beta_{k}E_{k}+\alpha_{k}}$,
instead of the usual Boltzmann weights $\,\omega_{B}\propto e^{-\beta E}$,
which enhances energetic tunnelings by improving the sampling of rare
configurations. 

However, those MUCA coefficients $\left\{ \beta_{k},\alpha_{k}\right\} $
are \textit{a priori} unknowns, so requiring an algorithm for their
determination before a production run can be pursued. A successful
technique to estimate MUCA weights is iterative \cite{BERG_MUCA}
and involves computing histograms $H_{muca}\left(E\right)$ of the
energies $E_{k}\in\left[E_{0},\ldots,E_{max}\right].$ Thus, in the
beginning, one sets $\omega_{muca}^{0}(E_{k})=1$ for all energies
$E_{k}$, which is used on an usual METROPOLIS simulation to collect
the initial $\left(n=0\right)$ histogram $H_{muca}^{0}\left(E\right).$
Thereafter, it is applied a recursive update scheme for the MUCA weights
$\omega_{muca}^{n+1}\left(E\right)=\omega_{muca}^{n}\left(E\right)/H_{muca}^{n}\left(E\right),$
before starting a new simulation with the lately estimated weights.
This process is repeated till convergence is detected by a flatness
check on the histograms. In the end, at a large-$n$ limit, the multicanonical
entropy is ensured to recover the microcanonical one as $S_{muca}\left(E\right)\rightarrow S\left(E\right)$.

The multicanonical implementation on PHAST employs accumulated error-weighted
histograms, so statistics for weight estimations improves for every
repetition, while convergence is attained by Berg's weight recursion\cite{BERG_MUCA}.
Additionally, its parallel estimator for MUCA weights --- initially
proposed by Zierenberg \textit{et. al.} \cite{Scaling_props_parallel_multican_algorithm}
to study spin-systems --- present scaling properties computationally
exceeding more traditional approaches \cite{MUCAREM_REMUCA}. This
is due to a simple and quite efficient update scheme, illustrated
on Figure 2. First, each processor (node) is initialized with the
same MUCA coefficients, but a particular random seed, then they perform
independent Markov Chain Monte Carlo (MCMC) simulations to accumulate
local energy histograms. After one iteration cycle those histograms
are combined on the master node, which updates and then broadcasts
the new MUCA weights, the cycle is repeated till convergence is achieved.

\begin{figure}
\begin{centering}
\includegraphics[clip,width=9cm]{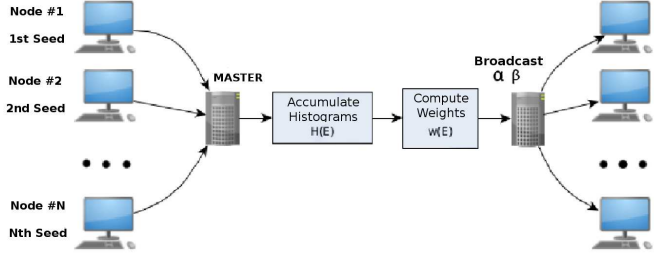}
\par\end{centering}

\caption{The iteration-cycle scheme for parallel multicanonical simulation:
N processors (nodes) are initialized with different random seeds and
so, evolve independent Markov Chains whose individual Histograms are
accumulated on $H\left(E\right)$ at each MUCA iteration to update
the weight $W\left(E\right).$ The resulting coefficients $\left(\alpha,\beta\right)$
on MUCA weights are then broadcasted.}
\end{figure}

\section{Software framework }

The PHAST package is constituted by the following modules
\begin{itemize}
\item SET\_INPUT: prepares input protein sequences from PDB files
\item PHAST: the main engine to run parallel multicanonical simulations
\item ANALYST: computes the microcanonical thermostatistics from MUCA simulations
\end{itemize}
It is not possible to give here a highly detailed description of all
program routines. Instead, we address the main features and data workflows
of aforementioned modules in the following subsections, which afterwards
is illustrated by some concrete study cases.

\subsection{SET\_INPUT}

A single text file named \texttt{\textit{ProteinSequences.txt}} is
to be provided as input to the simulation module. But while users
may manually encode each of its line entries, so describing protein
(or polymeric) sequences according to an AB-like lexicon (e.g. Eq.
\ref{Hydrophobicity_constant}), this can be alternatively generated
in a quicker manner. To this end, the SET\_INPUT module was designed
to facilitate the mapping of protein structures, downloaded from the
Protein Data Bank (PDB) \cite{PDB}, by using a built-in hidrophobicity
scale \cite{Roseman_scale} configured on file \texttt{\textit{hscale.h}}.

The aforesaid mapping is attained by invoking the setup module (\texttt{./set\_input.o})
and then typing the path to a (\texttt{\textit{.pdb }}or\texttt{\textit{
.seq}}) input file, whereas the software shall recognize among valid
PDB (\texttt{\textit{.pdb}}) or three-letter code (\texttt{\textit{.seq}})
sequences. Still, by successively calling this module with different
file sequences, more lines are appended to \texttt{\textit{ProteinSequences.txt}}\textit{,}
so making it quite easy to prepare input files even for simulations
of multiple-proteic systems.

\subsection{PHAST}

This is the main module of the package, here multicanonical algorithms
are implemented to simulate the coarse-grained models of Section 2.
In principle, PHAST just needs to read a \texttt{\textit{ProteinSequences.txt}}\textit{
}text file before starting any run. However, the software can be fully
reconfigured (before compilation) through editing its \texttt{\textit{.h}}
files, whose content is
\begin{itemize}
\item \texttt{vars.h}: the main configuration file that sets multicanonical
parameters. This includes, but is not limited to, the quantity and
size of energy binnings, the number of MCMC sweeps and MUCA iterations,
the definition of common blocks, the maximal size and number of proteins
as well as of the residue families allowed in a simulation.
\item \texttt{set\_hamiltonian.h}: enables to tune the coupling constants
specifying a particular Hamiltonian from Eq.(\ref{H_one_protein})
and Eq.(\ref{H_multi_protein}).
\item \texttt{box\_size.h}: defines the size of the steric spherical container
for proteins and some geometric cutoffs of polymer backbones.
\item \texttt{set\_randvec\_size.h}: random numbers may be optionally generated
as large vectors in GPUs, the size can be optimally adjusted to fit
memory caches.
\end{itemize}
Still, the main module is composed by the following briefly summarized
routines
\begin{itemize}
\item \texttt{MAIN.}F: is the principal routine of the simulation module,
contains all MPI communication structures and callings of I/O and
MUCA engine functions.
\item \texttt{HAMILTONIAN.}f: given a polymeric configuration this routine
computes the Hamiltonian defined by Eq.(\ref{H_one_protein}) and
Eq.(\ref{H_multi_protein}) with the constants adjusted at \texttt{set\_hamiltonian.h.}
\item \texttt{MONTE\_CARLO.}f: this is a wrap routine that employs the Metropolis
algorithm, with generalized MUCA weights, to perform MCMC stochastic
evolutions. Additionally, it updates the local energy histograms at
every simulation step.
\item \texttt{GEOMETRY.}f: the simulation code uses inner coordinates, as
bond-lengths and angles between beads, to evolve individual chains.
This routine allows to convert them to (external) cartesian coordinates
for computing inter-chain distances.
\item \texttt{\textbf{UPDATE\_ENGINE.}}\textbf{f:} this function stochastically
updates all bead-positions of the chains contained in the 3d-spherical
box defined at \texttt{box\_size.h}. It employs a mix of optimized
updating algorithms, to know, \textit{spherical-caps}: that changes
internal angles of each monomer along the backbone, \textit{pivotings}:
rotates a chain over a pivot direction, \textit{bond-length changes}:
by dilations or shrinks around the equilibrium value $r_{0}$ (if
$\kappa_{F}\neq0$), and \textit{Galilean-group updates}: i.e. rotations
plus translations of chains around the box center of mass. More details
on references in\textbf{\textit{\footnotesize{} }}\cite{h-c-r-IAPP_agg_PRE_RBFrigori}\textbf{\textit{\footnotesize .}}{\footnotesize \par}
\item \texttt{SET\_HF-MATRIX.}f: this routine permits to manually set the
interaction coefficients $C_{LJ}\left(\sigma_{i},\sigma_{j}\right)$
and $C_{F}\left(\sigma_{i},\sigma_{j}\right)$ of the Hamiltonian
in Eq.(\ref{H_one_protein}). It shall be modified accordingly to
\texttt{hscale.h,} which allows for consistent generalizations of
the AB-model lexicon such as \cite{3-letter_AB_model}.
\item \texttt{SET\_MUCA.}f: a bundle of auxiliary functions needed to set
MUCA simulations.
\item \texttt{SET\_PROT-SEQ.}f: reads in proteins from the AB-sequence file
\texttt{\textit{ProteinSequences.txt}}\textit{.}
\item \texttt{SET-SAVE\_SIM.}f: reads or writes to disk the simulation files,
as protein configurations \texttt{\textit{aAcidPosRxyz\_{*}.pdb}}
with extremal energies, produced during MCMC evolutions. 
\item \texttt{SIM\_IO-FILES.}f: reads or writes to disk the inner control
files of the simulation.
\item \texttt{RANDOM.f}: the wrap function for random number generation
on CPUs or GPUs.
\item \texttt{cudaRandom.cu}: enables random number generation on GPUs.
\end{itemize}
It also shall be noted that as long as the simulations progress a
serie of output files is automatically saved to disk. While files
as \texttt{\textit{Protein3DPositions.{*}}}\textit{, }\texttt{\textit{aAcid3DPositions.{*}}}\textit{
}and\textit{ }\texttt{\textit{last.con}}\textit{ }are just intended
for inner control of simulations, being specially needed when restarting
simulations, there are also many other output files with \texttt{\textit{{*}.dat}}
or \texttt{\textit{{*}.pdb}} extensions. As an example \texttt{\textit{beta\_n.dat,
alpha\_n.dat, g\_n.dat, }}and\texttt{\textit{ hist\_n.dat}}\textit{
}respectively denote files containing data from the \texttt{\textit{n}}\texttt{-th}
MUCA repetition, as the coefficients $\left\{ \beta_{k},\alpha_{k}\right\} _{n}$,
the density of states $g_{n}\left(E_{k}\right)$ and histograms $H_{n}\left(E_{k}\right)$
--- which are read by the ANALYST module afterwards --- as well as
the aforementioned \texttt{\textit{aAcidPosRxyz\_{*}.pdb}} files.

\subsection{ANALYST}

This module is the kernel of the package for microcanonical thermostatistical
analysis. A set of files, originally\textit{ }saved by the simulation
module \textit{--- i.e. }\texttt{\textit{beta\_{*}.dat, alpha\_{*}.dat,
}}and\texttt{\textit{ last.con}}\textit{ ---} is read and semi-automatically
processed. As an outcome neat \texttt{xmGrace} \cite{Grace} files
are produced according to template files, named as \texttt{\textit{template\_{*}.agr}}.
Some internal parameters shall be set before compilation, depending
on the intended output features, to know
\begin{itemize}
\item \texttt{icontrol}: set intensive units of energy, as \textquotedblleft{}energy
per volume\textquotedblright{}, on the outputs.
\item \texttt{iAGR:} defines if the thermodynamical observables of Eqs.(\ref{Boltzmann_formula})--(\ref{Free_energy})
are written to individual \texttt{.agr} files, or combined into a
single graph otherwise.
\item \texttt{Tol\_Beta \& Tol\_L: }respectively set (i) the numerical tolerance
among three successive $\beta$ values and (ii) the minimal latent
heat to be searched for during a Maxwell construction \cite{Gross_microcanonical_thermostatistics}
on $\beta\left(\varepsilon\right)\times\varepsilon$.
\end{itemize}
In particular, when the ANALYST module is called (\texttt{./analyst.o})
the \texttt{\textit{last.con}} file is read and supplies the software
with parameters (set at \texttt{vars.h}) used on that former MUCA
simulation. Then, the user is offered the option to type the configuration-ranges
of the iteration files to be analyzed. Finally, the following additional
information is requested to
\begin{itemize}
\item ``\textsf{Enter the number of monomers}'': necessary to output data
as intensive quantities.
\item ``\textsf{Enter the finite difference step-size, dE: {[}0.01,9.9{]}}'':
the energy step \textsf{dE} is used to compute the finite-difference
derivatives taken on $S\left(E\right).$
\item ``\textsf{Enter the $\beta$ value to compute Free Energy, or 0 for
Maxwell construction}'': during first-order like structural phase
transitions, the S-bends appear on microcanonical caloric curves.
So users may wish to provide a specific $\beta_{c}$ to compute the
Eq.(\ref{Free_energy}), or let the program find this pseudo-critical
value by a Maxwell construction.
\end{itemize}

\section{Example runs}

As case studies we have investigated two proteic systems by using
the AB-model limit%
\footnote{Note that by taking $\kappa_{F}=0$ the FENE interaction is disabled
and so all bond-lengths stay fixed to $r_{0}$ (e.g. $r_{0}=1$),
therefore the bead-stick limit of AB-model is recovered. %
} \cite{AB_model} of the Hamiltonian Eq.(\ref{H_multi_protein}).
Such modelling does not allows for prediction of protein structures,
but may provide an useful method to learn about some general thermodynamical
mechanisms underlying biological structural phase transitions \cite{Microcan_thermostatistics_coarse-grained_proteins}.
For instance, this idea was successfully employed to compute aggregation
propensities of peptides \cite{h-c-r-IAPP_agg_PRE_RBFrigori}.

\begin{verbatim}
a typical "set_hamiltonian.h" file
!-------------------------------------------------------------        
! Hamiltonian Parameters                                             
!-------------------------------------------------------------        
! Set c_KF, c_KC, c_KLJ_intra or c_KLJ_inter to 0.0d0 to   
! Shutdown any of that interactions  
!-------------------------------------------------------------
! FENE parameters: c_KF*R^2*ln(1-[(r-r0)/R]^2)         
!-------------------------------------------------------------         
r0 = 1.00d0     ! Distance of equilibrium between beads         
R  = 1.2d0      ! Scale parameter         
c_KF = 0.0d0    ! FENE coupling constant          
!-------------------------------------------------------------        
! Curvature parameters: c_KC*(1-cos*(Phi))
!-------------------------------------------------------------         
c_KC = 0.25d0     ! Curvature coupling constant 
!-------------------------------------------------------------  
! Non-bonded interaction parameters: c_KLJ*(1/R^12-sigma/R^6)      
!-------------------------------------------------------------         
c_KLJ_intra = 4.00d0   ! Intra-protein LJ coupling constant         
c_KLJ_inter = 4.00d0   ! Inter-protein LJ coupling constant
\end{verbatim}

Therefore, while keeping in mind the limitations of our coarse-grained
approach, we choose some natural amyloidogenic protein sequences as
inspiration to generate aggregation-prone heteropolymeric chains through
their hydrophobic mappings. So, we first simulate the aggregation
of highly amyloidogenic segments of the Amyloid $\beta$ protein,
associated to the Alzheimer disease --- i.e. the $A\beta_{25-33}$
segment, PDB:\textbf{2LFM } --- a system also applied for assessing
the parallel scaling (Speedup) of our code. Additionally, we have
simulated the aggregation of an heterogeneous molecular system, composed
by segments of $A\beta_{25-33}$ and Amylin isoforms (related to the
type 2 Diabetes Mellitus, see \cite{h-c-r-IAPP_agg_PRE_RBFrigori})
--- i.e. the human $hIAPP_{20-29}$ PDB:\textbf{2KB8} and rat $rIAPP_{20-29}$
PDB:\textbf{2KJ7} --- to test for catalytic effects of crowding \cite{IAPP_under_crowding}
and \textit{cross-seeding }\cite{Protein-protein_interaction} on
the onset of aggregation that induces such degenerative diseases.

\subsection{Aggregation of $A\beta_{25-33}$ segments: parallel speedup}

To simulate the aggregation of $A\beta_{25-33}$ segments, we initially
downloaded the 2LFM file from PDB \textbf{\cite{PDB} }and SET\_INPUT
converted it to an AB-sequence. This sequence was then duly cut, cloned
and used to prepare an input file to simulate the molecular aggregation
(dimerization), as can be checked in the next data sample

\begin{verbatim}
ProteinSequences.txt
BBBBBAAAB 
BBBBBAAAB
\end{verbatim}The box-radius $\left(R\right)$ was set to $R=40$ for these simulations
and, overall statistics collected by the multiple processors was fixed
to $10^{6}$ MCMC sweeps per MUCA iterations ($1.2\times10^{6}$ for
128 processors), in a total of $10^{3}$ repetitions. 

Before performing the data analysis of the resulting MUCA coefficients
$\left\{ \beta_{k},\alpha_{k}\right\} _{n},$ it was necessary to
ensure they had reasonably converged after some minimal number of
MUCA iterations $\left(N_{conv.}\right).$ To do so, we inspected
the flatness of energy histograms $H_{n}\left(E_{k}\right)$ in the
the full range of MUCA iterations, i.e. $n\in[1,1000].$ As an adequacy
criteria we imposed that the Coefficient of Variation $\left(\mathrm{c_{Var}}=\nicefrac{\sigma_{H}}{\mu_{H}}\right),$
defined as the ratio of the standard deviation $\left(\sigma_{H}\right)$
over the average value of a histogram $\left(\mu_{H}\right)$, should
be smaller than $5\%$. Additionally, local peaks on any supposedly
flat histogram were to be no larger than $20\%$ of its average value.
Considering both such constraints, we found acceptable convergence
(i.e. enough histogram flatness) was established for data in the iteration
range $n\in\left[N_{conv.}>175,1000\right]$, see Figure 3.

\begin{figure}
\begin{centering}
\includegraphics[clip,width=8.25cm]{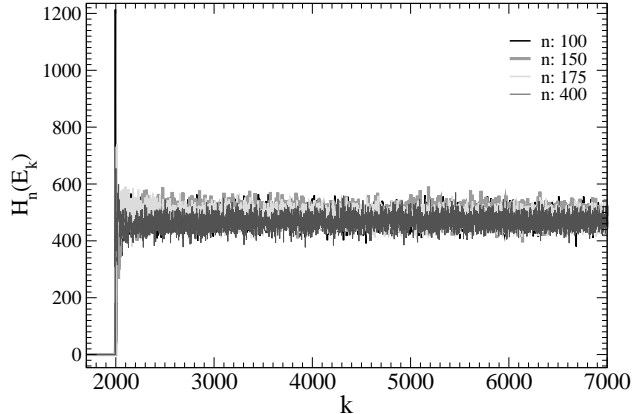}
\par\end{centering}

\caption{Representative energy histograms $H_{n}\left(E_{k}\right)$ employed
for flatness checks, which also allows for identifying the convergence
of MUCA coefficients $\left\{ \beta_{k},\alpha_{k}\right\} _{n}$
during simulations of $A\beta_{25-33}$ segments. For histograms in
the iteration range $n\in(175,1000]$ we observe acceptable Coefficient
of Variation $\left(\mathrm{c_{Var}}\leq5\%\right),$ see text. }
\end{figure}

Then, the ANALYST module was invoked to accomplish the thermostatistical
data analysis. To ensure a robust estimation of the error bars, various
starting $\left(N_{S}\geq N_{conv.}\right)$ and ending $\left(N_{E}\leq1000\right)$
points in the iteration range $n\in\left[N_{S},N_{E}\right]$ of MUCA
coefficients $\left\{ \beta_{k},\alpha_{k}\right\} _{n}$ were tested,
as well as the size of data-blocks $\triangle N$ (see Binning method,
\cite{Bergs_Book}) to be grouped in between. We also required the
numerical stability and inter-compatibility (within the statistical
errors) when computing the pseudo-critical inverse temperatures $\beta_{c},$
which was particularly important to improve the curve matchings during
data collapses (Figure 4). Despite tedious, this procedure also enabled
controlled checks of autocorrelation time effects over the statistical
error bars.

\begin{figure}
\begin{centering}
\includegraphics[clip,width=8.25cm]{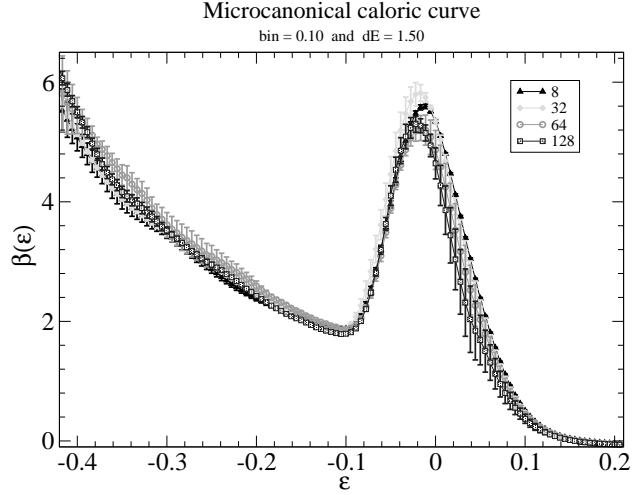}
\par\end{centering}

\caption{The data collapse of representative curves, of the inverse temperature
$\beta\left(\varepsilon\right)\times\varepsilon,$ illustrates the
equivalence on thermodynamical results from simulations employing
up to 128 processors.}
\end{figure}

\begin{figure}
\begin{centering}
\includegraphics[clip,width=8.5cm]{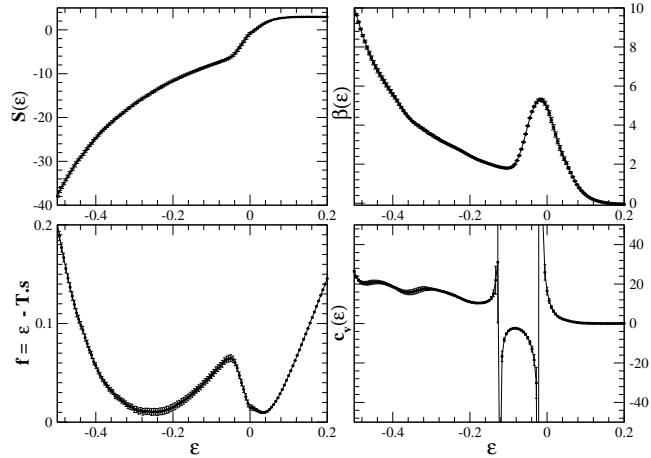}
\par\end{centering}

\caption{The microcanonical thermodynamic observables as seen from ANALYST
module for the aggregation of $A\beta_{25-33}$ fragments, to know:
the entropy $S\left(\varepsilon\right)\times\varepsilon$, caloric
curve $\beta\left(\varepsilon\right)\times\varepsilon,$ the (shifted)
free energy $f\left(=\varepsilon-\beta_{c}^{-1}\cdot S\right)\times\varepsilon,$
and specific heat $c_{v}\left(\varepsilon\right)\times\varepsilon.$}
\end{figure}

\begin{figure}
\begin{centering}
\includegraphics[clip,width=8.5cm]{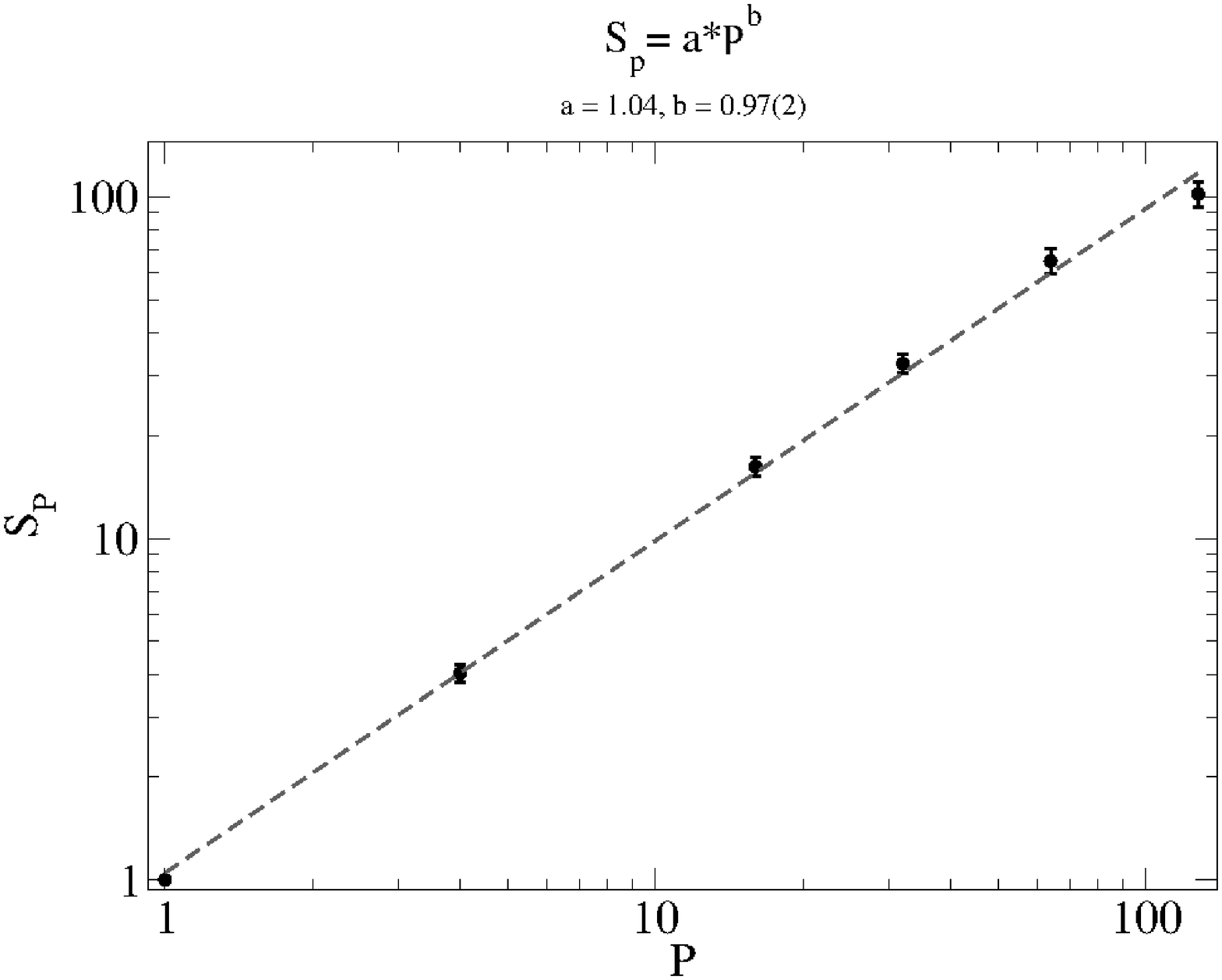}
\par\end{centering}

\caption{The parallel Speedup obtained for Multicanonical simulations present
almost linear scaling up to hundred processors, even at numerically
difficult first-order like phase transitions.}
\end{figure}

The physical consistency of our parallel results was validated by
the data collapse of multiple simulations, performed with increasing
number of CPUs (up to 128 processors) at the IBM P750 machine \cite{CENAPAD},
see a representative output on Figure 4 for parameters $\left[N_{S},N_{E}\right]=\left[200,1000\right]$
and $\triangle N=50$. The general microcanonical behavior of the
system was finally checked by an automatized plotting of the thermostatistical
observables saved to an\texttt{ .agr} file, as can be seen from Figure
5 (where $\left[N_{S},N_{E}\right]=\left[700,1000\right]$ and $\triangle N=25$).
It is observed a first-order like structural phase transition, characterized
by a reentrant on the microcanonical entropy $S\left(\varepsilon\right)$,
which induces an S-bend on the caloric curve $\beta\times\varepsilon$
and also a region with negative values of the microcanonical specific
heat $C_{v}\left(\varepsilon\right).$ See \cite{Gross_microcanonical_thermostatistics}
for details and a comprehensive review on this formalism. It is also
worth to note that such transitions are signaled by the presence of
energetic barriers on the free energy, a feature we have previously
studied on similar bioinspired molecular systems \cite{Microcan_thermostatistics_coarse-grained_proteins,h-c-r-IAPP_agg_PRE_RBFrigori}. 

To conclude, the wall times for simulations presenting equivalent
thermodynamical results --- which was ensured by performing a previous
data collapse --- were employed to analyze the parallel speedup of
our code, seen on Figure 6. So we plotted the parallel Speedup factor
$S_{p}=\nicefrac{t_{1}}{t_{p}},$ defined as the ratio of the serial
$t_{1}$ to the parallel $t_{p}$ times spent to accomplish the convergence
of MUCA weights. A power-law model $S_{p}=a\cdot P^{b}$ was proposed
for regression, and so established an almost linear%
\footnote{However, for a fixed total simulation statistics the error-bar sizes
are enlarged by increasing the number of processors. This probably
indicates an eventual speedup saturation, possibly connected to the
correlation times, as already observed in \cite{Scaling_props_parallel_multican_algorithm,Speedup_MUCAP_FENE}. %
} scaling $b=0.97\left(2\right)$ with the number of processors $P$.
This corroborates the nice scaling properties of the parallel MUCA
algorithm found on mild first-order like phase transitions \cite{Scaling_props_parallel_multican_algorithm,Speedup_MUCAP_FENE}.

\subsection{Cross-seeding of $A\beta_{25-33}$ and $IAPP{}_{20-29}$ isoforms}

Recent clinical studies have suggested that DM2 can be a catalyst
factor to the emergence of some neurodegenerative diseases, as the
AD \cite{AD_DM2_clinical-evidence}. A microscopic molecular explanation,
from \textit{in vitro }studies \cite{Heterogeneous_fibrils_AB-IAPP},
points to the \textit{cross-seeding} of $A\beta$ and $IAPP$ as a
key factor for the appearance of heterogeneous amyloid fibrils with
considerable cytotoxicity. This phenomena would be somehow mimicked
by the aggregation of bioinspired protein-like heteropolymers, investigated
by the methodology previously applied to study the aggregation of
$A\beta_{25-33}.$ Therefore, an input file was prepared by editing
the AB-sequences obtained from SET\_INPUT, once it is feed with the
2LFM ($A\beta$) and 2KB8 (hIAPP) files downloaded from PDB website
\cite{PDB}. That is

\begin{verbatim}
ProteinSequences.txt
BBBABAAABB  
BBBABAAABB  
BBBBBAAAB  
BBBBBAAAB
\end{verbatim}also, an equivalent system for IAPP molecules of rats (PDB:\textbf{2KJ7)}
was built, as shown

\begin{verbatim}
ProteinSequences.txt
BBBABAAAAA  
BBBABAAAAA  
BBBBBAAAB  
BBBBBAAAB
\end{verbatim}

\begin{figure}
\begin{centering}
\includegraphics[clip,width=8.5cm]{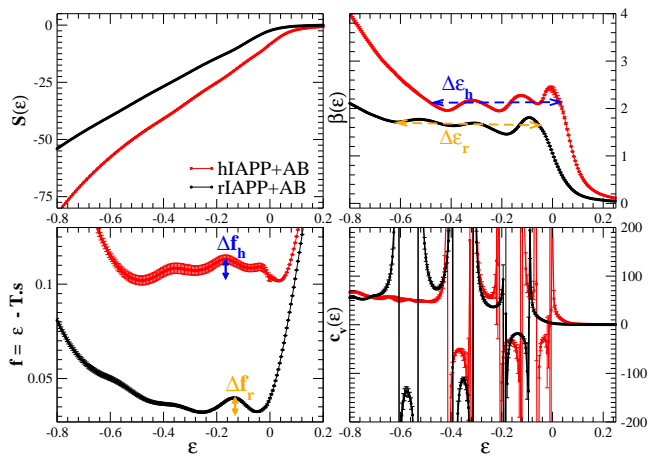}
\par\end{centering}

\caption{The microcanonical thermodynamic observables as seen from ANALYST
module for the aggregation of $A\beta_{25-33}$ fragments cross-seeded
by Amylin isoforms $hIAPP_{20-29}$ or $rIAPP_{20-29}$. To know:
the entropy $S\left(\varepsilon\right)\times\varepsilon$, caloric
curve $\beta\left(\varepsilon\right)\times\varepsilon$ including
the Maxwell construction, the free energies $f\left(=\varepsilon-\beta_{c}^{-1}\cdot S\right)\times\varepsilon,$
and the specific heat $c_{v}\left(\varepsilon\right)\times\varepsilon.$
While the energetic barrier $\triangle f_{h}=0.034(5)$ for the aggregation
of Human (h) molecular system is larger than the rodent (r) energetic
barrier $\triangle f_{r}=0.021(3),$ its groundstate is found at much
higher energies on the landscape. This explains the larger latent
heats for the rodent system $\triangle\varepsilon_{r}=0.584\left(5\right)$
{[}to be compared to $\triangle\varepsilon_{h}=0.513\left(5\right)${]},
and so its increased stability against aggregation. The energy axis
$\varepsilon=E/N$ is shown in intensive units.}
\end{figure}

\begin{figure}
\begin{centering}
\includegraphics[clip,width=9cm]{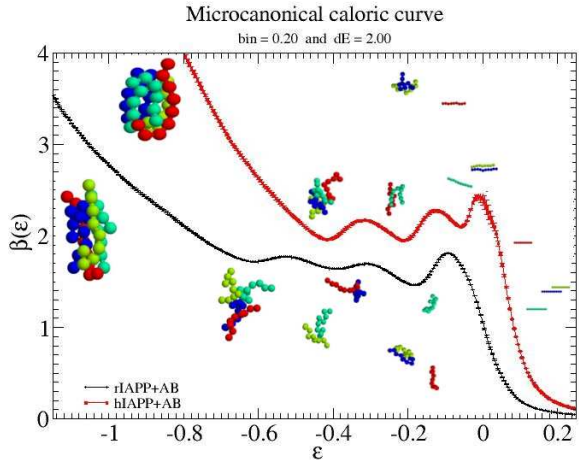}
\par\end{centering}

\caption{The microcanonical caloric curve $\beta\left(\varepsilon\right)\times\varepsilon$
with representative molecular configurations for the cross-seeding
of fragments of $A\beta_{25-33}$ and Amylin isoforms: $hIAPP_{20-29}$
or $rIAPP_{20-29}.$}
\end{figure}

These simulations were run on a considerably denser environment, once
the box-radius was set to $R=20.$ The 480 CPUs at \cite{SDumont}
shared a workload of $10^{7}$ MCMC sweeps per MUCA iteration, in
a total of $10^{3}$ MUCA repetitions. The obtained MUCA weights in
the iteration range $\left[N_{S},N_{E}\right]=\left[700,1000\right]$
were analyzed with the ANALYST module, while taking $\triangle N=25$.
Thermodynamical results are summarized for all microcanonical observables
of both systems on Figure 7, while details for representative structures
are exhibited on Figure 8. It is clearly observed that while the energetic
barrier for the aggregation of the human molecular system (i.e. $A\beta_{25-33}$
and $hIAPP{}_{20-29}$) $\triangle f_{h}=0.034(5)$ is slightly larger
than that of the rats system $\triangle f_{r}=0.021(3),$ the latter
transition happens at much deeper values of the free energy. This
fact justifies the larger latent heat found on the rodent system $\triangle\varepsilon_{r}=0.584\left(5\right)$
{[}to be compared to that of humans $\triangle\varepsilon_{h}=0.513\left(5\right)${]},
which explains thermodynamically its improved stability against aggregation.

Interesting enough, we also detected the presence of multiple backbendings
on $\beta\left(\varepsilon\right)\times\varepsilon$ (see Figure 8),
which are related to nucleation hierarchies, as broadly studied on
\cite{Multiple_backbendings}. Initially one observes the formation
of one hetero-oligomer made of one segment of $A\beta_{25-33}$ and
an Amylin isoform $(hIAPP_{20-29}$ or $rIAPP_{20-29}),$ whose thermodynamical
signature emerges by the first bump on the caloric curves. Following,
another hetero-oligomer is similarly built and, then finally the aggregation
process is completed by a global collapse of both hetero-oligomers
which results on a globular structure. It deserves to be noted that
those aggregates are clearly not amyloidic ones, as no cross-beta
structure is formed, which would require specific interactions for
their stabilization a feature absent in our Hamiltonian. However,
these protein-like aggregation transitions exhibit hierarchical nucleations
as would also be expected by cooperative \textit{cross-seeding }mechanisms
on real proteins.

Therefore, some intriguing conclusions may be addressed from our microcanonical
analysis, mainly regarding the stability of Amylin isoforms when \textit{cross-seeded}
by $A\beta$ segments. For example, a biotechnologically designed
protein named Pramlintide \cite{Pramlintide} --- originally inspired
on PDB:2KJ7 --- has been recently used as an adjuvant to treat DM2
in humans, curiously such protein becomes identical to $rIAPP$ when
mapped by the hydrophobic scale \cite{Roseman_scale} to the AB-model.
Thus, our simulations indicate that Pramlintide could also be employed
for damping the aggregation of heterogeneous amyloidogenic systems,
as ones made by $A\beta$ and $IAPP$ isoforms. This result is corroborated
not only by aggregation propensities previously computed from the
AB-model \cite{h-c-r-IAPP_agg_PRE_RBFrigori}, but also by studies
of similar systems \cite{IAPP_under_crowding,Protein-protein_interaction}
using all-atom force fields.

\section{Conclusions}

We have presented PHAST, a package for simulating coarse-grained models
of proteins and bioinspired polymers in the multicanonical ensemble.
Despite of its simple structure, the software comprises a reconfigurable
force field which allows for an excellent parallel speedup. Additionally,
its modules enable users to quickly map proteins downloaded from PDB
to an inner AB-lexicon --- or its generalizations as \cite{3-letter_AB_model}
--- as well as to automatically compute most of systems microcanonical
thermostatistics. Still, the program codes are open and free, which
shall motivate students and researchers to readily adapt them to their
specific purposes, as modelling many-body protein-protein interactions
or the realistic effects of crowding on polymeric systems. We plan
to gradually improve the built-in force fields, besides moving their
calculations to GPUs, and implement faster evolution algorithms as
hybrid MCMC \cite{HMC}.

\section*{Acknowledgements}

This is a long-term project whose author has benefited from enlightening
discussions with Leandro G. Rizzi, Lieverton H. Queiroz, Mathias S.
Costa, Mikhael C. Chrum, and Nelson A. Alves. The Brazilian laboratories
CENAPAD-SP and LNCC-RJ are acknowledged by providing machine time
and helpful technical support, respectively on their IBM P750 \cite{CENAPAD}
and Santos Dumont \cite{SDumont} supercomputing facilities.

\end{document}